\documentclass[aps,pre,twocolumn,amsmath,amssymb,amsfonts]{revtex4-1}
\pdfoutput=1
\usepackage{graphicx}
\usepackage{dcolumn}
\usepackage{bm}
\usepackage{amsmath}
\usepackage{amsfonts}
\usepackage{amsgen}
\usepackage{amsbsy}

\usepackage{footnote}
\usepackage[titletoc,title]{appendix}
\usepackage{color}
\usepackage[normalem]{ulem} 

\newcommand\D{\mathrm{d}}
\newcommand\E[1]{\mathrm{e}^{#1}}
\newcommand\Dif[2]{\frac{\mathrm{d}{#1}}{\mathrm{d}{#2}}}

\newcommand\diag{\mathrm{diag}}

\begin{document}

\title{Quartic polynomial approximation for fluctuations of separation of trajectories in chaos and correlation dimension}

\author{Itzhak Fouxon$^{1,2}$}\email{itzhak8@gmail.com} \author{Siim Ainsaar$^1$}\email{ainsaar@gmail.com}\author{Jaan Kalda$^1$}\email{kalda@ioc.ee}
\affiliation{$^1$Department of Cybernetics, School of Science at Tallinn University of Technology, Tallinn, 12 618, Estonia}
\affiliation{$^2$Department of Computational Science and Engineering, Yonsei University, Seoul 120-749, South Korea}

\begin{abstract}

We consider the cumulant generating function of the logarithm of the distance between two infinitesimally close trajectories of a chaotic system. Its long-time behavior is given by the
generalized Lyapunov exponent $\gamma(k)$ providing the logarithmic growth rate of the $k-$th moment of the distance. The Legendre transform of $\gamma(k)$ is a large deviations function that gives the probability of rare fluctuations where the logarithmic rate of change of the distance is much larger or much smaller than the mean rate defining the first Lyapunov exponent. The only non-trivial zero of $\gamma(k)$ is at minus the correlation dimension of the attractor which for incompressible flows reduces to the space dimension. We describe here general properties constraining the form of $\gamma(k)$ and the Gallavotti-Cohen type relations that hold when there is symmetry under time-reversal. This demands studying joint growth rates of infinitesimal distances and volumes. We demonstrate that quartic polynomial approximation for $\gamma(k)$ does not violate the Marcinkiewicz theorem on invalidity of polynomial form for the generating function.
We propose that this quartic approximation will fit many experimental situations, not having the effective time-reversibility and the short correlation time properties of the quadratic Grassberger-Procaccia estimates. We take the existing $\gamma(k)$ for turbulent channel flow and demonstrate that the quartic fit is nearly perfect. The violation of time-reversibility for the Lagrangian trajectories of the incompressible Navier-Stokes turbulence below the viscous scale is considered. We demonstrate how the fit can be used for finding the correlation dimensions of strange attractors via easily measurable quantities. We provide a simple formula via the Lyapunov exponents, holding in quadratic approximation, and describe the construction of the quartic approximation. A different approximation scheme for finding the correlation dimension from expansion in the flow compressibility is also provided.

\end{abstract}

\maketitle

\section{Introduction}

Positivity of the Lyapunov exponent $\lambda_1$ is the most widely used definition of chaos \cite{yb,dor,k,review}. Infinitesimally close trajectories separate in time exponentially with the growth exponent $\lambda_1$ which is the same for (almost) all trajectories in the limit of infinite time \cite{oseledets}. However care is needed in the usage of this result. The moments  $\langle r^k(t)\rangle$  of the inter-particle distances $r(t)$, averaged with respect to the initial position of the pair, behave at large times as $\exp(\gamma(k)t)$. Then $r\sim \exp(\lambda_1 t)$ would imply $\gamma(k)=\lambda_1 k$ however this behavior is forbidden.
Indeed, $\gamma(k)$ must vanish at $k$ equal to minus the correlation dimension of the attractor \cite{do} (that for incompressible flows reduces to the space dimension \cite{dl}) which rules out a linear $\gamma(k)$, cf. \cite{fuji,pario,crisa}. The non-linear behavior, known as intermittency, originates in non-uniform convergence of $\lambda(t)\equiv t^{-1}\ln (r(t)/r(0))$ to $\lambda_1$. However large the time is, there are spatial regions such that the pairs issuing from them have $\lambda(t)$ values strongly differing from $\lambda_1$. The volume fraction of the initial positions having $\lambda(t)=\lambda\neq\lambda_1$ decays with time exponentially. The decay exponent $H(\lambda)$ is called the rate or large deviations function and is the Legendre transform of $\gamma(k)$. The ratio of $r(t)\sim \exp(\lambda t)$ for these rare trajectories to the most probable growth $\exp(\lambda_1 t)$ can also exponentially grow in time. As a result $\langle r^k(t)\rangle$ is determined by the exponentially decaying in time fraction of trajectories for any $k$.

The function $\gamma(k)$ describes the asymptotic behavior of the cumulant generating function \cite{lu} of the logarithm of the distance $\rho(t)\equiv \ln (r(t)/r(0))$. It
contains significantly more information on the system than the Lyapunov exponent given by $\gamma'(0)=\lambda_1$. Since it provides a certain logarithmic growth rate of the distance then it is called the generalized Lyapunov exponent. This property of a chaotic system have been studied for a long time, see e. g. \cite{fuji,pario,crisa}, however recently there appeared new measurements where the chaotic system is formed by the motion of the fluid particles resolved below the smoothness (viscous \cite{frisch}) scale of the Navier-Stokes (NS) turbulence. Thus $\gamma(k)$ or $H(\lambda)$ were obtained for the homogeneous \cite{mj} and channel \cite{Bagheri,mj12} turbulent flows. The dependence on the Reynolds number in the case of homogeneous turbulence was considered in \cite{bc} who also studied $\gamma(k)$ for the chaotic motion of inertial particles (\cite{bc,mj,mj12} used a somewhat different definition of $\gamma(k)$ which coincides with ours at $k>-1$. We use the more traditional definition \cite{pario,crisa} used also in \cite{Bagheri}, see later). These measurements produced $\gamma(k)$ not describable by the parabolic approximation of Grassberger and Procaccia \cite{quadratic}. This spurred our interest in finding an efficient fitting form that could describe the observations. In this paper we introduce the general properties that constrain $\gamma(k)$ and demonstrate that the quartic approximation for $\gamma(k)$ is consistent.  The question of consistency is raised by the Marcinkiewicz theorem that tells that the cumulant generating function cannot be a polynomial of a degree larger than two \cite{lu}. Using the data, kindly provided by the authors of \cite{Bagheri}, we demonstrate that the quartic approximation fits the data almost perfectly.

The generalized Lyapunov exponent has had a growing number of applications coming from the fluid mechanics. It was used for demonstrating the growth of small fluctuations of magnetic field $\bm B$ in an incompressible flow of conducting fluid with negligible magnetic resistivity \cite{dl}. In this case the magnetic field lines behave as the material lines of the fluid \cite{ll8} and $\lim_{t\to\infty}t^{-1} \ln |B(t)/B(0)|=\lambda_1$ where $\bm B(t)$ is measured on the trajectory of the fluid particle. The growth then holds for generic flows where $\gamma(k)$ does not vanish identically. Indeed, $\gamma(k)$ is a convex function as seen readily from H\"older's inequality (the cumulant generating function is convex \cite{lu}). It has two zeros: a trivial zero at $k=0$ and a non-trivial zero at $k=-d$ where $d$ is the space dimension  \cite{dl}. Hence $\gamma'(0)$, that gives the field growth exponent $\lambda_1$, must be positive. Moreover the non-linearity of $\gamma(k)$ helps to stress the role of intermittency: $\gamma(2)$ ($\neq 2\lambda_2$) provides the growth rate of the magnetic energy which determines the growing relevance of the field's back reaction on the flow.

The above proof of positivity of $\lambda_1$ generalizes to compressible flows. The motion of particles in these flows is a dissipative dynamical system so that typically the trajectories at large times asymptote a multifractal attractor \cite{dor,ruelle}. It can be demonstrated that the non-trivial zero of $\gamma(k)$ is located at minus the correlation dimension \cite{procor,hp} of this attractor $D$, see \cite{do,krzysztof}. The correlation dimension can be defined via the scaling exponent $r^{D-d}$ of the probability of finding a pair of trajectories separated by a small $r$ after a long evolution time. The dimension $D$, as a fractal dimension, is enclosed between zero and $d$. In generic cases $D$ is strictly smaller than $d$ so that $r^{D-d}$ diverges in correspondence with the singularity of the steady state density supported on a multifractal set \cite{dor}. Thus the second zero of $\gamma(k)$ is still negative and $\gamma'(0)=\lambda_1$ must be positive. We conclude that the Lyapunov exponent of a generic dissipative dynamical system is positive.

The correlation dimension $D$, providing the non-trivial zero of $\gamma(k)$, is one of the most applied of the fractal dimensions \cite{procor}. It determines the collision kernel of particles transported by fluids where an effective flow of the particles, different from the fluid flow that is assumed to be incompressible, can be introduced. An example is provided by weakly inertial particles \cite{maxey}. In the limit of negligible inertia these particles are tracers whose motion coincides with that of the fluid particles. However the finite inertia causes a centrifugal effect which repels the particles from the vortices. This effect is captured by the formula for the particle velocity that is given by the local flow plus a correction term describing the repulsion from the vortices. The correction is minus the particle's reaction time multiplied by the local acceleration of the fluid particles. Thus the particle's velocity, despite differing from the local flow, is still a function of the particle's position i.e. the flow of particles can be introduced. This flow is already compressible since the correction has a non-zero divergence, providing a finite, albeit small, particles' flow compressibility. The compressibility results in the particles' accumulation on a mulitfractal attractor, located below the viscous scale \cite{frisch} where the flow and its vortices are smooth. This comparatively small region of scales of turbulence is of applied value since small particles collide at those scales as in the case of rain formation by water droplets, see e. g. \cite{FFS}. Similar cases where the particles's velocity is given by the sum of the local flow and a compressible local correction are provided by the motile phytoplankton cells \cite{2013,fls,2019}, phoretic particles \cite{phor1,phor2} and fine bubbles \cite{fineb}. Somewhat different situation is provided by water droplets sedimenting in warm clouds where usually gravitational acceleration is larger than the turbulent one. In the limit of a much stronger gravity even strongly inertial particles form a smooth flow. However in this case the particle's velocity depends on turbulence non-locally in space and time \cite{2015}. In this case $\rho(t)$ obeys the Langevin equation and $\gamma(k)$ is quadratic as in \cite{quadratic}. Yet another case of a dissipative system is provided by tracers confined to the surface. The motion can be driven by the underwater turbulence \cite{uw1,uw2} (where \cite{uw2} obtained the large deviations function) or the surface wave turbulence \cite{sw1,sw2,sw3}. In all these cases particles' collisions are of high interest and their rate depends on the value of $D$ characterizing how often the particles' distances approach the interaction distance. The correlation dimension can also be considered for the inertial particles in turbulence for parameters where there is no smooth flow in space, using the flow in the six-dimensional phase space \cite{bc}. In this case the relation between the collision rate and $D$ demands future study. The provided examples demonstrate that there is plenty of applications for $\gamma(k)$ and dynamical systems framework in the fluid mechanics.

In this work we describe universal properties of the generalized Lyapunov exponent that hold irrespective of the details of the flow. We also describe a generalized version of $\gamma(k)$ that involves one more argument
describing the joint growth rates of products of distances $r(t)$ and infinitesimal volumes. For compressible flows, in contrast to $\gamma(k)$, the two-argument exponent obeys a closed constraint. The information on $\gamma(k)$ can then be obtained as a marginal distribution.

The quadratic approximation for $\gamma(k)$, that was introduced in \cite{quadratic} for studying the correlation dimension, is often too restrictive, see also \cite{pb}. Indeed, $\gamma'(0)=\lambda_1$ and $\gamma(-D)=0$ uniquely fixes $\gamma(k)$ as $\lambda_1k(k+D)/D$. This form implies many constraints that would be often violated strongly. For instance for incompressible flow, where $D=d$, the quadratic approximation necessitates the equality of $\lambda_1$ and minus $d-$th Lyapunov exponent $\lambda_d$ (the exponents are defined via the logarithmic growth rate of hypersurfaces of different integer dimensions. For instance in the physical dimension three $\lambda_1$ is the logarithmic growth rate of infinitesimal line elements, $\lambda_1+\lambda_2$ - of infinitesimal area elements and $\sum_{i=1}^3\lambda_i$ is the logarithmic growth rate of infinitesimal volumes).   The equality $\lambda_d=-\lambda_1$ is in fact true if the incompressible flow is also time-reversible. However when the flow is not time-reversible and/or compressible the equality generally breaks down and $\lambda_1\neq -\lambda_d$. For instance the Lagrangian trajectories of the three-dimensional incompressible NS turbulence, which is not time-reversible, obey $\lambda_3\approx -5\lambda_1/4$, see \cite{review} and references therein. Thus \cite{quadratic} considered the possibility that higher-order corrections might be necessary. The quartic polynomial approximation, proposed here, addresses this necessity. It is determined by three readily measurable phenomenological parameters. It seems that this approximation works in the NS case. We prove this for the channel turbulence of \cite{Bagheri} and observe that the fit would probably also work for the measurements of \cite{mj} which provide $H(\lambda)$ that seemingly can be fit with a simple function. Since many of the physical examples provided above have weakly compressible flows then we also introduce an approximation scheme for $D$ where the flow compressibility is considered as an expansion parameter.

Some relations of this work appeared previously in the PhD Thesis of one of the authors \cite{thesis} however were never published. The main progress achieved here in this direction is relaxation of the restrictive assumption of isotropy and description of implications of time-reversal symmetry.


\section{Intermittency of chaotic separation and generalized Lyapunov exponent} \label{interm}

We consider evolution of the distance $\bm r(t)$ between two infinitesimally close trajectories $\bm x_1(t)$ and $\bm x_1(t)+\bm r(t)$ of a chaotic $d-$dimensional system $\dot {\bm x}=\bm v(t, \bm x(t))$. The flows
$\bm v(t, \bm x)$ of interest here are either time-independent or stationary random flows. In this Section the flow can be incompressible or compressible and it is assumed to be smooth below a certain scale $\eta$ (viscous scale in the NS case). We introduce $\sigma_{ik}(t)=\nabla_k v_i(t, \bm x_1(t))$ so that for $r\ll \eta$ the evolution is governed by $\dot{\bm r}=\sigma\bm r$. The solution of this equation can be written as $\bm r(t)=r_0\exp\left(\rho(t)\right){\hat n}(t)$ where the unit vector ${\hat n}(t)$ obeys,
\begin{eqnarray}&&\!\!\!\!\!\!\!\!\!
\Dif{{\hat n}}{t}=\sigma {\hat n}-\xi {\hat n},\ \
\rho(t)=\int_0^t \xi(t')dt',\ \ \xi(t)\equiv {\hat n}\sigma{\hat n}. \label{reps}
\end{eqnarray}
The most famous property of the evolution of $r(t)$ is the existence of the trajectory-independent limit
\begin{eqnarray}&&\!\!\!\!\!\!\!\!\!
\lim_{t\to\infty}\frac{\rho(t, \bm x, {\hat n}(0))}{t}=\lambda_1, \label{first}
\end{eqnarray}
which defines the first Lyapunov exponent $\lambda_1$. Similar representation holds for $-\lambda_d$ where $\sigma$ must be changed to $-\sigma^T$, with $T$ standing for transpose, see Appendix of \cite{fb}. The equation tells that the limit exists and it does not depend on the initial position, $\bm x_1(0)=\bm x$, and the initial orientation of the pair, ${\hat n(0)}$, despite the fact that it could. The independence from $\bm x$ is seen by observing that $\rho(t)/t$ is a time-average, similar to that appearing in the ergodic theorem, see Eq.~(\ref{reps}).
The independence of ${\hat r}={\hat n}(0)$ can be seen by introducing the Jacobi matrix $W_{ik}(t, \bm x)=\nabla_kq_i(t,\bm x)$.
This matrix is defined by taking derivatives of positions $\bm q(t, \bm x)$ of the system trajectories at time $t$ with respect to their initial position $\bm x$,
\begin{eqnarray}&&\!\!\!\!\!\!\!\!\!
\partial_t \bm q(t, \bm x)=\bm v(t, \bm q(t, \bm x)),\ \ \bm q(0, \bm x)=\bm x.
\end{eqnarray}
Thus $\bm q(t, \bm x)$ are the Lagrangian trajectories of the fluid formed by the continuum of the trajectories. The Oseledec theorem states that \cite{oseledets},
\begin{eqnarray}&&\!\!\!\!\!\!\!\!\!
\lim_{t\to\infty}\frac{\ln W^T(t, \bm x)W(t, \bm x)}{2t}
\nonumber\\&&\!\!\!\!\!\!\!\!\!
=N^T(\bm x) \diag[\lambda_1, \lambda_2,\ldots,\lambda_d] N(\bm x),\label{osl}
\end{eqnarray}
where $N(\bm x)$ is an orthogonal matrix and $\diag[\lambda_1, \lambda_2,\ldots,\lambda_d]$ is the diagonal matrix whose values $\lambda_i$, arranged in non-increasing order $\lambda_{i+1}\geq \lambda_i$, define the Lyapunov exponents. The exponents are independent of $\bm x$ for almost all $\bm x$. We observe that $\bm r(t)$ can be written with the help of $W(t)$ as $\bm r(t)=r_0W(t, \bm x){\hat r}$. Thus
$2\rho(t)=\ln {\hat r}  W^T(t, \bm x)W(t, \bm x) {\hat r}$. We find from Eq.~(\ref{osl}) that Eq.~(\ref{first}) holds for all initial directions ${\hat r}$ that have non-zero projection on the vector with components $N_{1i}$ provided $\lambda_1>\lambda_2$, see \cite{fp} for detailed study.

The convergence of the limit in Eq.~(\ref{first}) is strongly non-uniform with respect to $\bm x$. For most $\bm x$ at long time scales, approximation $\rho(t)\approx \lambda_1 t$ holds. Thus if we randomly seed a pair of close trajectories at $t=0$, then for most of them we would find $r(t)\sim \exp(\lambda_1 t)$.
However it would be wrong to conclude that the moments $\left\langle r^k(t)\right\rangle$ behave as $\exp(k\lambda_1 t)$. In fact there is no $k$ for which $\left\langle r^k(t)\right\rangle\sim \exp(k\lambda_1 t)$ holds strictly (for small $k$ it is true approximately). Here we use for averaging, designated by angular brackets, the usual space averaging over $\bm x$ (other type of averaging which is used often employs the natural measure \cite{ruelle}: the two averages would usually coincide, see \cite{gawedzki} and below). The growth is described by the generalized Lyapunov exponent $\gamma(k)$,
\begin{eqnarray}&&\!\!\!\!\!\!\!\!\!
\gamma(k)\equiv \lim_{t\to\infty}\frac{\ln \left\langle \exp(k\rho(t))\right\rangle}{t}= \lim_{t\to\infty}\frac{\ln \left\langle |W{\hat r}|^k\right\rangle}{t}. \label{de}
\end{eqnarray}
Thus $\gamma(k)$ is a limit of rescaled cumulant generating function $\ln \left\langle \exp(k\rho(t))\right\rangle$ of the random variable $\rho(t)$ (taken at imaginary argument) \cite{lu}. Convexity of the
cumulant generating functions implies that $\gamma(k)$ is also convex. The name generalized Lyapunov exponent was used in \cite{mj} for objects obtained by using $\rho_i$ or their linear combinations instead of $\rho$ in Eq.~(\ref{de}). Our definition is more traditional and allows the use of analytical properties of $\gamma(k)$, cf. the Introduction. The formula for $\gamma(k)$ in terms of the exponents of \cite{mj} is provided below.

We demonstrate that the $t\to\infty$ limit in Eq.~(\ref{de}) exists and is independent of the initial orientation ${\hat r}$. We use the decomposition $W=R(t)\Lambda(t)N(t)$ where $R(t)$ and $N(t)$ are orthogonal matrices and $\Lambda=\diag[\exp(\rho_1(t)), \exp(\rho_2(t)),\ldots,\exp(\rho_d(t))]$, see Appendix of \cite{fl} and also \cite{fb,review,cge}. The quantities $\rho_i(t)$ have a simple interpretation:
a small ball of radius $\epsilon$ is transformed into an ellipsoid whose axes' lengths are $\epsilon \exp(\rho_i(t))$. The Oseledec theorem asserts the existence of finite limits $\lim_{t\to\infty}\rho_i(t)/t=\lambda_i$ and $\lim_{t\to\infty}N(t)=N$, see Eq.~(\ref{osl}). We find at large times,
\begin{eqnarray}&&\!\!\!\!\!\!\!\!\!
 \left\langle|W(t){\hat r}|^k\right\rangle\sim \left\langle|\Lambda(t) N{\hat r}|^k\right\rangle
 \nonumber\\&&\!\!\!\!\!\!\!\!\!
 =\left\langle[\exp(2 \rho_1(t))m_1^2+..+\exp(2 \rho_d(t))m_d^2]^{k/2}\right\rangle, \label{moments}
\end{eqnarray}
where $m_i= (N{\hat r})_i$. Here the time independence of $N(t)$ is the consequence of $\rho_i(t)\gg \rho_{i+1}(t)$, see \cite{fl} (the Oseledec theorem cannot be employed because we consider the object determined by large deviations from the behavior described by the theorem). The average in Eq.~(\ref{moments}) is determined by an optimal fluctuation for which all $\rho_i$ scale linearly with $t$. This can be seen from the large deviations form of the probability density function (PDF) $P(\{\rho_i\}, t)$ of $\rho_i(t)$ that obeys,
\begin{eqnarray}&&\!\!\!\!\!\!\!\!\!
P(\{\rho_i\}, t)\sim \exp\left(-t{\tilde H}\left(\frac{\rho_1}{t},\ldots,\frac{\rho_d}{t}\right)\right), \label{ld}
\end{eqnarray}
where ${\tilde H}(\bm x)$ is the convex large deviations function. This function has an unique minimum equal to zero at $x_i=\lambda_i$ so that at $t\to\infty$ it reproduces $\lim_{t\to\infty}\rho_i(t)/t=\lambda_i$. Similarly one can find the central limit theorem for $(\rho_i(t)-\lambda_i t)/\sqrt{t}$, see \cite{fb} and also \cite{review,gawedzki}. The conditions under which Eq.~(\ref{ld}) holds are that $\sigma_{ik}(t)$ have a
finite correlation time and the spectrum of Lyapunov exponents is non-degenerate, $\lambda_i>\lambda_{i+1}$. These conditions are assumed to hold. Then the average in Eq.~(\ref{moments}) is,
\begin{eqnarray}&&\!\!\!\!\!\!\!\!\!
\left\langle |W(t){\hat r}|^k\right\rangle \sim \int\exp\left(-t{\tilde H}\left(\frac{\rho_1}{t},\ldots,\frac{\rho_d}{t}\right)\right)
\nonumber\\&&\!\!\!\!\!\!\!\!\!
[\exp(2 \rho_1(t))m_1^2+..+\exp(2 \rho_d(t))m_d^2]^{k/2} \prod_{i=1}^d \D\rho_i.
\end{eqnarray}
This integral is determined at large times by the saddle-point. The saddle-point values of all $\rho_i$ scale linearly with time. Thus in the leading order the resulting average does not depend on the constants $m_i^2$ bounded between zero and one as long as $m_i\neq 0$. We conclude that $\gamma(k)$ is independent
of ${\hat r}$ except for ${\hat r}$ that obey $(N{\hat r})_i=0$ for some $i$, cf. similar condition for the validity of Eq.~(\ref{first}) above. These vectors have zero measure on the sphere and will be of no interest here. Thus we can use an equivalent definition of $\gamma(k)$,
\begin{eqnarray}&&\!\!\!\!\!\!\!\!\!
\gamma(k)= \lim_{t\to\infty}\frac{\ln \int \left\langle |W(t){\hat r}|^k\right\rangle \D{\hat r}}{t}=\lim_{t\to\infty}\frac{1}{t}
\\&&\!\!\!\!\!\!\!\!\!
\ln \int  \left\langle \left[\exp(2 \rho_1(t)){\hat r}_1^2+..+\exp(2 \rho_d(t)){\hat r}_d^2\right]^{k/2} \right\rangle \D{\hat r}. \nonumber
\end{eqnarray}
This form of the definition converges faster in time and is useful both experimentally and theoretically. Applying H\"older's inequality to moments of $r(t)$ we confirm that $\gamma(k)$ is a convex function.

Cumulant expansion theorem \cite{ma} provides a series representation for $\gamma(k)$. In accord with the assumption that $\sigma_{ik}(t)$ has a finite correlation time, we assume that $\xi(t)$ in Eq.~(\ref{reps}) has a finite correlation time $\tau_c$. We stress that $\tau_c$ does not necessarily coincide with the correlation time of $\sigma_{ik}(t)$ because the angular degree of freedom ${\hat n}(t)$ can change the structure of temporal correlations, as in the example of anisotropic Kraichnan model \cite{cge}. The cumulants of $\rho(t)$ are proportional to $t$ at $t\gg \tau_c$ giving,
\begin{eqnarray}&&\!\!\!\!\!\!\!\!\!
\gamma(k)= k\lambda_1+\frac{k^2\Delta}{2} +\frac{k^3}{3!}\int_{-\infty}^{\infty} \langle \xi(0)\xi(t_1)\xi(t_2)\rangle_c \D t_1\D t_2
\nonumber\\&&\!\!\!\!\!\!\!\!\!+\ldots;\ \ \ \
\Delta\equiv \int_{-\infty}^{\infty} \langle \xi(0)\xi(t)\rangle_c \D t,\label{gamma}
\end{eqnarray}
where we used $\langle \rho(t)\rangle=\lambda_1 t$. Here the dots stand for higher order cumulants and the subscript $c$ stands for cumulant. An interesting question for future work is to determine the radius of convergence of this series and when $\gamma(k)$ is an entire function.

We observe from Eq.~(\ref{gamma}) that $\gamma'(0)=\lambda_1$ however $\gamma(k)\neq k\lambda_1$ so that
$\left\langle r^k(t)\right\rangle\propto \exp(k\lambda_1 t)$ does not hold. This corresponds to strongly intermittent growth of $r(t)$. This growth is described by the large deviations function $H(\lambda)$ described in the Introduction, see e.g. \cite{review} and references therein for the large deviations formalism. That function gives asymptotic form of the PDF $P(\rho, t)$ of $\rho(t)$ at large times,
\begin{eqnarray}&&\!\!\!\!\!\!\!\!\!
P(\rho, t)\sim \exp\left(-tH\left(\frac{\rho}{t}\right)\right). \label{las}
\end{eqnarray}
We see that $H(\lambda)$ with $\lambda=\rho/t$ is the Legendre transform of $\gamma(k)$,
\begin{eqnarray}&&\!\!\!\!\!\!\!\!\!
\left\langle \exp(k\rho(t))\right\rangle\!\sim\! \int \!\!\exp\left(k\rho\!-\!tH\left(\frac{\rho}{t}\right)\right)\D\rho
\label{trabs}\\&&\!\!\!\!\!\!\!\!\!
\sim \exp\left(t\max_{\lambda}[k\lambda\!-\!H\left(\lambda\right)]\right),\ \ \gamma(k)\!=\!\max_{\lambda}[k\lambda\!-\!H\left(\lambda\right)],\nonumber
\end{eqnarray}
where the integral is obtained asymptotically at large time scales. Thus, $H(\lambda)$ is also convex. Setting $k=0$ in $\gamma(k)=\max_{\lambda}[k\lambda-H\left(\lambda\right)]$ and using $\gamma(0)=0$ demonstrates that $H(\lambda)$ is a non-negative function. The (unique) minimum of zero is attained at $\lambda=\lambda_1$ as seen by considering the Legendre transform formula $\gamma(k)=k\lambda(k)-H(\lambda(k))$, where $k=H'(\lambda(k))$.
Setting $k=0$ in $\gamma'(k)=\lambda(k)+\left[k-H'(\lambda(k))\right]\lambda'(k)$ gives $\lambda(0)=\lambda_1$. Thus the PDF $\exp(-tH(\lambda))$ of $\lambda(t)\equiv \rho(t)/t$, see Eq.~(\ref{las}), becomes $\delta(\lambda-\lambda_1)$ at $t\to\infty$.
This reproduces Eq.~(\ref{first}) that holds with probability one. We also find that the PDF $P(\tau, t)\sim \exp(-tH(\lambda_1+\tau/\sqrt{t}))$ of
the variable $\tau(t)\equiv \sqrt{t}[\lambda(t)-\lambda_1]$ obeys the central limit theorem,
\begin{eqnarray}&&\!\!\!\!\!\!\!\!\!
\lim_{t\to\infty} P(\tau, t)=\frac{1}{\sqrt{2\pi \Delta}}\exp\left(-\frac{\tau^2}{2\Delta}\right),
\end{eqnarray}
where $\Delta$ defined in Eq.~(\ref{gamma}) equals $1/H''(\lambda_1)$. This result describes typical deviations of the finite-time Lyapunov exponent $\lambda(t)$ from its infinite-time limit $\lambda_1$,
cf. \cite{gawedzki}.

The formulas above provide a quantitative description of the intermittency described in the Introduction.
We see from Eq.~(\ref{trabs}) that the moment of order $k$ is determined by rare fluctuations with $\lambda(k)\neq \lambda_1$ where $\lambda(k)$ gives the maximum to $k\lambda-H\left(\lambda\right)$. The probability of such fluctuations for which $r(t)\sim r(0)\exp(\lambda(k)t)$ is exponentially small, and is given by $\exp\left(-tH\left(\lambda(k)\right)\right)$. However, they increase the observable $r^k\sim r^k(0)\exp(k\lambda(k)t)$ so much that these rare events end up determining the value of this moment. Consider as an example $\langle r^k(t)\rangle$ with large negative $k$. For typical events with $\lambda=\lambda_1$ the observable is exponentially small, and hence, the contribution of the typical events into the average is negligible. Meanwhile, the average grows with time exponentially, and the growth is dominated by exponentially rare events for which the initial perturbation decays and the distance between the trajectories contracts exponentially. The probability of these events vanishes exponentially with time, however they increase the observable $r^k(t)$ also exponentially. The increase is so large that it is these rare contraction events that determine the average at $k<1-d$, as we demonstrate in the Sec. \ref{thro}. Thus the moment is determined by exponentially small fraction of pairs of particles which disappear at a very fast rate. This fact complicates the measurements and is known as the Lagrangian intermittency.

We remark that the above exponential time dependence of $\langle r^k(t)\rangle$ holds indefinitely for the moments that are determined by the contracting fluctuations. In contrast, for moments that are determined by the events with growing $r(t)$, the exponential time dependence is cut off at the times when $r(t)$ becomes of the order of the smoothness scale of the flow. Beyond these times, the Taylor approximation for the velocity difference of the diverging trajectories breaks down and a study relying on the detailed structure of the large-scale flow is needed.

Finally we provide the counterpart of the definition of the Lyapunov exponents for the time-reversed flow $-\bm v(t, \bm x)$. We observe that we can also consider the evolution of a small ball of radius $\epsilon$ backwards in time starting from time zero. The axes of the ellipsoid are in this case $\epsilon\exp(\rho_i(t))$ where $t<0$ and $\rho_i(t)\geq \rho_{i+1}(t)$ is still assumed. Then the limits,
\begin{eqnarray}&&\!\!\!\!\!\!\!\!\!
\lambda_i^-=\lim_{t\to - \infty}\frac{\rho_i(t, \bm x)}{|t|},
\end{eqnarray}
define the Lyapunov exponents of the time-reversed flow $\lambda_i^-$. For instance, $\lambda_1^-$ gives the backward in time logarithmic divergence rate of infinitesimally close trajectories, cf. \cite{oseledets,gawedzki}. For incompressible flow, by the reversal of time direction, $\lambda^-_i=-\lambda_{d-i+1}$ and statistics of $\rho_i(-|t|)$ coincides with the statistics of $-\rho_{d-i+1}(|t|)$. For compressible flow, the relation between
the forward and backward in time exponents becomes non-trivial because of non-conservation of volumes, see \cite{arxiv} and also \cite{gawedzki}. One finds that probability density $P(\rho_i, -|t|)$ of $\rho_i(-|t|)$ is,
\begin{eqnarray}&&\!\!\!\!\!\!\!\!\!
P(\rho_i, -|t|)\!\sim\! \exp\left(\!-\!\sum_{i=1}^d \rho_i\!-\!|t|{\tilde H}\left(-\frac{\rho_d}{|t|},\ldots, -\frac{\rho_1}{|t|}\right)\right), \label{ld1}
\end{eqnarray}
where ${\tilde H}(\bm x)$ is the same function as in Eq.~(\ref{ld}) (this formula is of Gallavotti-Cohen type \cite{gawedzki}). The normalization of this PDF is the consequence of the conservation of the total volume of the flow \cite{arxiv},
\begin{eqnarray}&&\!\!\!\!\!\!\!\!\!
\langle J\rangle=\int J(t, \bm x)\frac{\D\bm x}{\Omega}=\left\langle \exp\left(-\sum_{i=1}^d \rho_i\right)\right\rangle=1, \label{o}
\end{eqnarray}
where we introduced the Jacobian $J(t, \bm x)=\det W(t, \bm x)$. Finally we introduce the generalized Lyapunov exponent of the time-reversed flow,
\begin{eqnarray}&&\!\!\!\!\!\!\!\!\!
\gamma^-(k)= \lim_{t\to-\infty}\frac{\ln \int \left\langle |W(t){\hat r}|^k\right\rangle \D{\hat r}}{|t|}.
\end{eqnarray}
The properties of the time-reversed flow will be useful in the study of the forward in time quantities.

\section{Generalized sum of Lyapunov exponents} \label{s}

In this Section we introduce a generalized sum of Lyapunov exponents, $s(p)$, which is very useful for the study of compressible flows \cite{thesis}. Using the interpretation of the compressible flow as a dissipative dynamical system, $s(p)$ describes fluctuations of the average entropy production $-\int_0^t w(t', \bm q(t', \bm x))dt'/t$ over a finite time interval $t$ \cite{dor}, where $w(t, \bm x)\equiv \nabla\cdot\bm v(t, \bm x)$.
At large $t$ this quantity is positive with probability close to one, however there is a finite probability of negative entropy production providing a finite time violation of the second law of thermodynamics. For time-reversible statistics the Gallavotti-Cohen relation determines the ratio of probabilities of a given entropy production and minus this value, see \cite{dor} and references therein.

We observe that evolution of infinitesimal volumes of the fluid is described by the Jacobian of the Lagrangian map $\det W(t, \bm x)$, where $W_{ik}(t, \bm x)$ was defined in the previous Section. The equation on infinitesimal volumes of the fluid \cite{batc} gives,
\begin{eqnarray}&&
\frac{\partial \ln J(t, \bm x)}{\partial t}=w(t, \bm q(t, \bm x)).
\end{eqnarray}
Thus the logarithmic rate of the growth of infinitesimal volumes obeys the ergodic theorem \cite{dor},
\begin{eqnarray}&&\!\!\!\!\!\!\!\!\!
\lim_{t\to\infty}\frac{\ln J(t, \bm x)}{t}\!=\!\lim_{t\to\infty}\!\frac{1}{t}\int_0^t \!\!\!w(t', \bm q(t', \bm x))\D t'\!=\!\sum_{i=1}^d\lambda_i.
\end{eqnarray}
The limit defines the sum of Lyapunov exponents $\lambda_i$ which is readily seen to be consistent with the definitions of $\lambda_i$ in the previous Section. Similar limit can be considered for the time-reversed evolution,
\begin{eqnarray}&& \!\!\!\!\!\!\!\!\!
\lim_{t\to -\infty}\!\!\!\frac{\ln J(t, \bm x)}{|t|}\!=\!\!\!\lim_{t\to-\infty}\!\!\int_0^{t}\!\!\!\! w(t', \bm q(t', \bm x))\frac{\D t'}{|t|}\!=\!\sum_{i=1}^d \lambda_i^-,
\end{eqnarray}
cf. \cite{cod}. The limits in the sums hold for almost every $\bm x$, except for the possible exception of $\bm x$ with zero total volume. This reservation is significant here since it can be readily seen that the points $\bm x$ at which the limit $\lim_{t\to -\infty}\ln J(t, \bm x)/|t|$ differs from $\sum_{i=1}^d \lambda_i^-$ contain all the mass of the system supporting the singular steady state density (the natural measure) \cite{fz}. Similar fact holds for $\lim_{t\to -\infty}\ln J(t, \bm x)/t$ where the points at which the limit is not $\sum_{i=1}^d\lambda_i$ support the steady state density of the time reversed flow, cf. the repeller \cite{dor}.

In contrast with the rest of the Lyapunov exponents, the sums of Lyapunov exponents can be simply represented in terms of the flow. The sums can be written as time integrals of the different time correlation functions of the flow divergence \cite{ff,cod},
\begin{eqnarray}&&\!\!\!\!\!\!\!
\sum_{i=1}^3\!\lambda_i\!=\!-\!\!\int_0^{\infty}\!\!\!\!\!\!\langle w(0)w(t)\rangle \D t,\ \ \sum_{i=1}^3\!\lambda_i^-\!=\!-\!\!\int_{-\infty}^0\!\!\!\!\!\!\langle w(0)w(t)\rangle \D t,
\nonumber\\&&\!\!\!\!\!\!\! \langle w(0)w(t)\rangle=\int w(0, \bm x)w [t, \bm q(t, \bm x)]\D\bm x.\label{sums01}
\end{eqnarray}
Here $\langle w(0)w(t)\rangle$ is generally not even a function of $t$ since spatial averaging does not correspond to the average over the steady state density. It is true generally \cite{ruella,ff,cod} that
$\sum_{i=1}^3\!\lambda_i\leq 0$ and $\sum_{i=1}^3\!\lambda_i^-\leq 0$. When the flow is generic, as will be assumed below, the integrals of $\langle w(0)w(t)\rangle$ above are non-zero and both sums
of the Lyapunov exponents are negative. We observe a useful representation,
\begin{eqnarray}&&\!\!\!\!\!\!\!
\sum_{i=1}^3\!\lambda_i^-\!=\!\lim_{t\to-\infty}\frac{\sum_{i=1}^d\rho_i(t)}{|t|} =-\lim_{t\to\infty}\frac{1}{t}  \label{repl}
\\&&\!\!\!\!\!\!\!
\int \sum_{i=1}^d\rho_i\exp\left(\!\sum_{i=1}^d \rho_i\!-\!t{\tilde H}\left(\frac{\rho_1}{t},\ldots, \frac{\rho_d}{t}\right)\right)\prod_{i=1}^d \D\rho_i,\nonumber
\end{eqnarray}
where we averaged the deterministic limit using Eq.~(\ref{ld1}) and changed variables $\rho_i\to -\rho_{d-i+1}$.

\begin{figure}
  \centerline{\includegraphics[width=\linewidth]{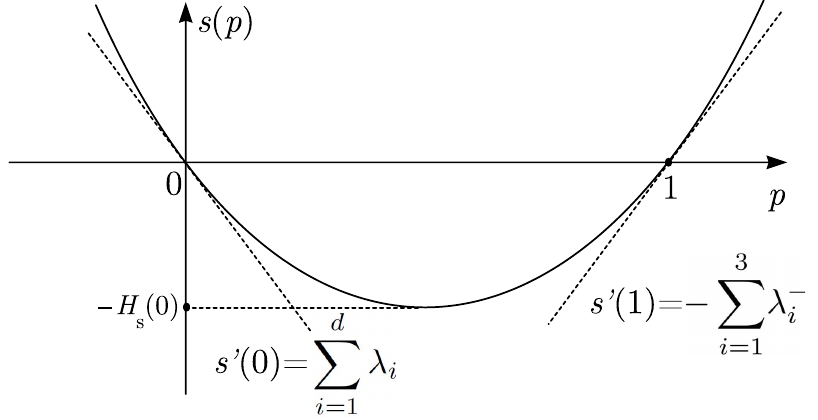}}
  \caption{Typical form of $s(p)$. The minimum of $s(p)$ gives the decay exponent of the probability of conservation of infinitesimal volume during time $t$.}
\label{fig:sp}
\end{figure}

We introduce the generalized sum of Lyapunov exponents $s(p)$,
\begin{eqnarray}&&\!\!\!\!\!\!\!\!\!
s(p)\!=\!\lim_{t\to\infty}\frac{\ln \left\langle J^p(t)\right\rangle}{t}\!=\!\lim_{t\to\infty}\frac{\ln \left\langle \exp\left(p\sum_{i=1}^d\rho_i(t)\right)\right\rangle}{t}.
\end{eqnarray}
We observe from the definition of $\sum_{i=1}^d\lambda_i$ and Eq.~(\ref{repl}) that,
\begin{eqnarray}&&\!\!\!\!\!\!\!
s'(0)\!=\!\sum_{i=1}^d\lambda_i<0,\ \ s'(1)=-\sum_{i=1}^3\!\lambda_i^->0.
\end{eqnarray}
We consider the large deviations function that gives the distribution $P(\nu, t)\sim \exp\left(-t H_s(\nu/t)\right)$ of $\nu=\sum_{i=1}^d\rho_i(t)$.
We have,
\begin{eqnarray}&&\!\!\!\!\!\!\!
s(p)=\lim_{t\to\infty}\frac{1}{t}\ln \int \exp\left(p\nu-tH_s\left(\frac{\nu}{t}\right)\right),
\end{eqnarray}
which implies alongside with $\omega=\nu/t$,
\begin{eqnarray}&&\!\!\!\!\!\!\!
s(p)=\max_{\omega}[p\omega-H_s(\omega)],\ \ H_s(\omega)=\max_{p}[p\omega-s(p)].
\end{eqnarray}
Thus the minimum of $s(p)$ equals to $-H_s(0)$, giving the probability $\exp(-tH_s(0))=\exp(t\min[s(p)])$ of "incompressible" events on which infinitesimal volumes are conserved with exponential accuracy (entropy is conserved). Using the fact that $s(p)$ is a convex function that vanishes at $p=0$ and $p=1$, see Eq.~(\ref{o}), we find that $s(p)$ has the general form given in Fig.~\ref{fig:sp}.

We can similarly study the generalized sum of Lyapunov exponents $s^-(p)$ for the time-reversed flow,
\begin{eqnarray}&&\!\!\!\!\!\!\!
s^-(p)=\lim_{t\to-\infty}\frac{\ln \left\langle J^p(t)\right\rangle}{|t|}=\lim_{t\to-\infty}\frac{1}{t}
\\&&\!\!\!\!\!\!\!
\ln \int\! \exp\left((1\!-\!p)\sum_{i=1}^d \rho_i\!-\!t{\tilde H}\left(\frac{\rho_1}{t},\ldots, \frac{\rho_d}{t}\right)\right)\prod_{i=1}^d \D\rho_i,\nonumber
\end{eqnarray}
where we used Eq.~(\ref{ld1}). Thus $s^-(p)$ is not an independent function, $s^-(p)=s(1-p)$. Thus for flow that obeys time-reversible statistics we have $s(p)=s(1-p)$ so that $s(p)$ is symmetric with respect to $p=1/2$ (and $\sum_{i=1}^d\lambda_i=\sum_{i=1}^3\!\lambda_i^-$).
For time-reversible statistics obeying the relation $s(p)=s(1-p)$, we have
\begin{eqnarray}&&\!\!\!\!\!\!\!
H_s(\omega)\!=\!\max_{p}[p\omega\!-\!s(1\!-\!p)]
\\&&\!\!\!\!\!\!\!
\!=\!\max_{p}[(1\!-\!p)(-\omega)\!+\!\omega\!-\!s(1\!-\!p)]=H_s(-\omega)+\omega.\nonumber
\end{eqnarray}
This relation tells that the PDF $P(\omega, t)$ of $\omega(t)=\nu(t)/t$ obeys $P(-\omega, t)/P(\omega, t)=\exp(\omega t)$. Observing that $\omega(t)$ is minus the average entropy production in time $t$, we recognize the more common form of the Gallavotti-Cohen relation than Eq.~(\ref{ld1}), cf. \cite{dor}.

\section{Generalized Lyapunov exponent from finite-time Lyapunov exponents}\label{thro}

The relation between the generalized Lyapunov exponent $\gamma(k)$ and the statistics of the finite-time Lyapunov exponents depends on $k$ non-trivially. For positive $k$, the moment $\langle r^k(t)\rangle$ is determined by the events where $r(t)$ grows such that  $\langle r^k(t)\rangle\sim \left\langle \exp(k\rho_1(t))\right\rangle$. However, considering a decrease of $k<0$, the events with contracting $r(t)$ become more relevant and as we reasoned in Sec. \ref{interm}, the moments with large negative $k$ would rather obey $\langle r^k(t)\rangle\sim \left\langle \exp(k\rho_d(t))\right\rangle$. In this Section we derive $\gamma(k)$ from the statistics of $\rho_i(t)$.
This formula appeared previously in the PhD Thesis \cite{thesis}.

The study demands joint distribution of the generalized Lyapunov exponents introduced previously,
\begin{eqnarray}&&\!\!\!\!\!\!\!
\gamma(k, p)=\lim_{t\to\infty} \frac{\ln \int\left \langle J^p(t)|W(t){\hat r}|^k\right\rangle \D{\hat r}}{t}
\\&&\!\!\!\!\!\!\!
=\lim_{t\to\infty} \frac{1}{t} \ln \left \langle\exp\left(\int_0^t \left(pw(t')+k\xi(t')\right)\D t'\right)\right\rangle. \nonumber
\end{eqnarray}
We recover $\gamma(k)$ as $\gamma(k, p=0)$ and $s(p)$ as $\gamma(k=0, p)$. The function
$\gamma(k, p)$ is convex so that its Hessian is a positive definite matrix.

We consider the average in the form,
\begin{eqnarray}&&\!\!\!\!\!\!\!
\int\left \langle J^p(t)|W(t){\hat r}|^k\right\rangle \D{\hat r}
\sim \left\langle J^p(t) \int\right.\left[\exp(2 \rho_1(t))x_1^2+\right.
\nonumber\\&&\!\!\!\!\!\!\!
..+\left.\exp(2 \rho_d(t))x_d^2\right]^{k/2} \left.\delta\left(\sqrt{1-\sum_{i=1}^d x_i^2}\right)\D\bm x\right\rangle.
\end{eqnarray}
Over most of the sphere, at large times, $\exp(2 \rho_1(t))$ term dominates the sum. However there are also domains dominated by $\exp(2 \rho_k(t))$ with $1<k\leq d$. These domains are relevant for certain ranges of negative $k$.
Next, notice that $1=\sum_{l=1}^d \sum_{i\neq l} \theta\left(|x_l|-\exp\left(\rho_i-\rho_l\right)|x_i|\right)$ where $\theta(x)$ is the step function; using this identity in the integrand, we obtain
\begin{eqnarray}&&\!\!\!\!\!\!\!
\E{\gamma(k, p)t}\!\sim \!\left\langle \E{p\sum_{i=1}^d\rho_i}\sum_{l=1}^d \E{k\rho_l}\prod_{i=1}^l \E{\rho_l-\rho_i}\right\rangle.
\end{eqnarray}
We observe that the ratio of $l-$th term to $(l+1)-$term is $\exp\left((k+l)(\rho_l-\rho_{l+1})\right)$. We find,
\begin{eqnarray}&&\!\!\!\!\!\!\!
\gamma(k, p)\!\sim \!\frac{1}{t}\ln \left\langle \E{p\sum_{i=1}^d\rho_i+k\rho_1}\right\rangle,\ \ k>-1;\label{throguh}\\&&\!\!\!\!\!\!\!
\gamma(k, p)\!\sim \!\frac{1}{t}\ln \left\langle \E{p\sum_{i=1}^d\rho_i+(k+1)\rho_2-\rho_1}\right\rangle,\ \ -2\!<\!k\!<\!-1; \nonumber\\&&\!\!\!\!\!\!\!
\ldots \ \gamma(k, p)\!\sim \!\frac{1}{t}\ln \left\langle \E{(p-1)\sum_{i=1}^d\rho_i+(k+d)\rho_d}\right\rangle,\ \ k<1-d.\nonumber
\end{eqnarray}
These relations are useful and can be used for deriving the properties of the exponent. We can derive the largest and smallest Lyapunov exponents of the flow and its time-reversal from $\gamma(k, p)$ as
\begin{eqnarray}&&\!\!\!\!\!\!\!\!\!\!
\frac{\partial\gamma(-d, 1)}{\partial k}\!= \!\lambda_d,\ \ \frac{\partial\gamma(-d, 2)}{\partial k}\!= \!-\lambda_1^-,\ \ \frac{\partial\gamma(0, 1)}{\partial k}\!= \!-\lambda_d^-, \label{derivatives}
\end{eqnarray}
where as in the previous Section, $\partial_k\gamma(0, 0)\!= \!\lambda_1$ and the sums $\sum_{i=1}^d\lambda_i$ and $\sum_{i=1}^d\lambda_i^-$ can be obtained from $\gamma(0, p)$.
We find from Eqs.~(\ref{throguh}) a remarkable identity,
\begin{eqnarray}&&\!\!\!\!\!\!\!\!\!
\gamma(0, p)=\gamma(-d, p+1).\label{identityp}
\end{eqnarray}
We observe that $\gamma(0, p)=s(p)$ vanishes at $p=0$ and $p=1$. We then find that $\gamma(-d, 1)=\gamma(-d, 2)=0$. In the case of incompressible flow this reduces to the known \cite{dl} equality $\gamma(-d)=0$.

The corresponding formulas for $\gamma^-(k)$ are obtained by using Eq.~(\ref{throguh}) with the PDF given by Eq.~(\ref{ld1}). By changing integration variables from $\rho_i$ to $-\rho_{d-i+1}$, we find that
\begin{eqnarray}&&\!\!\!\!\!\!\!\!\!
\gamma^-(k, p)\!\sim \!\frac{1}{t}\ln \left\langle \E{(1-p)\sum_{i=1}^d\rho_i-k\rho_d}\right\rangle,\ \ k>-1.
\end{eqnarray}
We see by comparison with Eq.~(\ref{throguh}) that $\gamma(k, p)=\gamma^-(-d-k, 2-p)$ for $k<1-d$. We find similarly that
\begin{eqnarray}&&\!\!\!\!\!\!\!\!\!
\gamma^-(k, p)\!\sim \!\frac{1}{t}\ln \left\langle \E{(1-p)\sum_{i=1}^d\rho_i-(k+1)\rho_{d-1}+\rho_d}\right\rangle,
\end{eqnarray}
for $-2<k<-1$. This demonstrates that $\gamma(k, p)=\gamma^-(-d-k, 2-p)$ also holds for $1-d<k<2-d$. Continuing in this manner, the equality can be proved for all $k$. The last equality of this type is found from
\begin{eqnarray}&&\!\!\!\!\!\!\!\!\!
\gamma^-(k, p)\!\sim \!\frac{1}{t}\ln \left\langle \E{(2-p)\sum_{i=1}^d\rho_i-(k+d)\rho_1}\right\rangle,\ \ k<1-d,\nonumber
\end{eqnarray}
demonstrating $\gamma(k, p)=\gamma^-(-d-k, 2-p)$ for $k>-1$. We find a useful identity,
\begin{eqnarray}&&\!\!\!\!\!\!\!\!\!
\gamma(k, p)=\gamma^-(-d-k, 2-p).
\end{eqnarray}
This relation can be used for the effective measurement of $\gamma(k, p)$ via moments of time-reversed flow. This could be simpler for the moments whose forward in time evolution is contraction demanding high resolution. We see from the equation above that for time-reversible statistics,
\begin{eqnarray}&&\!\!\!\!\!\!\!\!\!
\gamma(k, p)=\gamma(-d-k, 2-p). \label{rgtr}
\end{eqnarray}
Thus, incompressible time-reversible flow obeys $\gamma(k)=\gamma(-d-k)$ considered in detail in Sec.~\ref{incr}.
We observe that for time-reversible statistics $\gamma(0, p)=\gamma(0, 1-p)$, derived in the previous Section, and Eq.~(\ref{rgtr}) reproduce Eq.~(\ref{identityp}).

We consider implications of time-reversibility for the large deviations function ${\tilde H}$ that determines the joint PDF $P(\rho, \nu, t)$ of $\rho(t)$ and $\nu(t)$
\begin{eqnarray}&&\!\!\!\!\!\!\!\!\!
P(\rho, \nu, t)\sim \exp\left(-t{\tilde H}\left(\frac{\rho}{t}, \frac{\nu}{t}\right)\right).
\end{eqnarray}
This function is the Legendre transform of $\gamma(k, p)$,
\begin{eqnarray}&&\!\!\!\!\!\!\!\!\!
{\tilde H}\left(\lambda, \omega\right)=\max_{k, p}\left[\lambda k+\omega p-\gamma(k, p)\right].
\end{eqnarray}
We find the Gallavotti-Cohen type relation for time-reversible statistics using Eq.~(\ref{rgtr}),
\begin{eqnarray}&&\!\!\!\!\!\!\!\!\!
{\tilde H}\left(\lambda, \omega\right)
={\tilde H}\left(-\lambda, -\omega\right)-\lambda d+2\omega.
\label{agskl}
\end{eqnarray}
Reduction of relations of this Section in the case of incompressible flow will be considered in Sec.~\ref{incr}.

\section{Inequality on $\gamma(-d)$} \label{ineq}

Previously we reproduced from Eq.~(\ref{identityp}) the incompressible flow identity $\gamma(-d)=0$. In this Section we consider how  finite compressibility changes $\gamma(-d)$.
We demonstrate that $\gamma(-d)\geq 0$ where the equality holds only for incompressible flow. We use compressible version of an identity for integrals over a unit sphere used in \cite{dl} for the study of the incompressible case,
\begin{eqnarray}&&
J^{-1}\int |W^{-1}{\hat r}|^k \D S=\int \frac{\D S' }{|W{\hat r}'|^{d+k}}, \label{identity}
\end{eqnarray}
where ${\hat r}$ and ${\hat r}'$ are unit vectors. To prove this identity, we consider a transformation of the unit sphere ${\hat r'}=W{\hat r}/
|W{\hat r}|$ and the corresponding tranformation $\D S\to \D S'$
of the surface element. We note that $\D S'=\bm n\cdot \D\bm S''
/|W{\hat r}|^{d-1}$, where $W{\hat r}=|W{\hat r}|{\hat n}$ and $\D\bm S''$
is the transformation of the surface element ${\hat r}\D S$ under
${\hat r}\to W{\hat r}$. Consideration of the latter
tranformation of the volume element $(\D r{\hat r}, {\hat r}\D S)$ gives
$|W{\hat r}|\bm n\cdot \D\bm S''=J \D S$. Collecting the
above together, we have $J \D S=|W{\hat r}|^d \D S'$ which gives Eq.~(\ref{identity}).
By averaging this equation and using independence of the averages of ${\hat r}$ at large times, we find that
\begin{eqnarray}&&
\gamma(-d-k)=\lim_{t\to\infty}\frac{1}{t}\ln \left\langle J^{-1}(t)|W^{-1}(t){\hat r}|^k\right\rangle,
\end{eqnarray}
whose validity does not need isotropy of the flow statistics, cf. \cite{dl,review}. After setting $k=0$, this yields
\begin{eqnarray}&&
\gamma(-d)=\lim_{t\to\infty}\frac{1}{t}\ln\left\langle J^{-1}(t)\right\rangle=s(-1)> 0,
\end{eqnarray}
where we assume that the flow is generic so that $\sum_{i=1}^d\lambda_i$ is strictly negative. This inequality implies that the correlation dimension of the dynamics' attractor is smaller than the space dimension, i.e. the attractor is strange, see Sec. \ref{cor}.
We finally observe that after multiplying Eq.~(\ref{identity}) with $J^{p+1}(t)$ and averaging the result and setting $k=0$, Eq.~(\ref{identityp}) is reproduced.

\section{Incompressible flow: time-reversibility and its breakdown for the NS} \label{incr}

In this Section, we derive the properties of $\gamma(k)$ for incompressible flows where $\sum_{i=1}^d\rho_i=0$. We will consider the two and three dimensional cases, which have direct applications to fluid flows, in more detail.

\begin{figure}
  \centerline{\includegraphics[width=\linewidth]{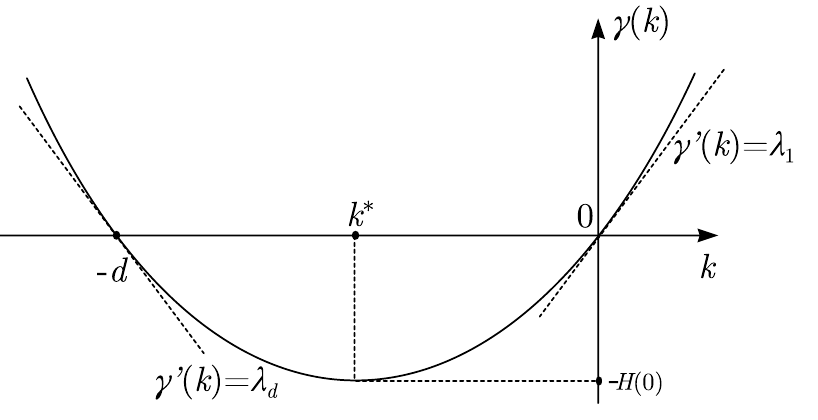}}
  \caption{Typical form of $\gamma(k)$ for an incompressible chaotic flow. The minimum of $\gamma(k)$ gives the decay exponent of the probability of conservation of distance between infinitesimally close trajectories during time $t$.}
\label{fig:gammak}
\end{figure}

We start from the observation that for $d-$dimensional incompressible flow, $\gamma(k)$ has the structure shown in Fig.~\ref{fig:gammak}.
This structure is fixed by the demands that $\gamma(k)$ is convex, vanishes at $k=0$ and $k=-d$ and obeys
$\gamma'(0)=\lambda_1$ and $\gamma'(-d)=\lambda_d$, see Eq.~(\ref{derivatives}). The unique minimum of $\gamma(k)$ holds at $k=k^*$ where $-d<k^*<0$. From $H(\lambda)=\max_k[k\lambda-\gamma(k)]$, we have that $\gamma(k^*)=-H(0)$. Thus the minimum value of $\gamma(k^*)$ gives the probability $\exp(-t H(0))=\exp(t\gamma(k^*))$ of conservation of the distance between two trajectories during time interval $t$. This conservation must hold with exponential accuracy so that $\rho(t)/t$ is small, cf. with a similar consideration in Sec.~\ref{s}. Finally, from the Legendre transform formula and $\gamma'(-d)=\lambda_d$ we find that $\lambda_d$ satisfies equality  $H(\lambda_d)=d|\lambda_d|$. Thus $\lambda_d$ can be determined from the plot of $H(\lambda)$ as the graph intersection with the straight line $-\lambda d$.

In the case of two dimensions, $\rho_i$ has one independent component. This strongly constrains the statistics so that Eqs.~(\ref{rgtr}) give $\gamma(k)=\gamma(-2-k)$. Thus, $\gamma(k)$ is symmetric with respect to $k=-1$ and $k^*=-1$. The large deviations function obeys $H(\lambda)=H(-\lambda)-2\lambda$, cf. Eq.~(\ref{agskl}).
Simple approximations for $\gamma(k)$ can be developed by truncating the Taylor series,
\begin{eqnarray}&&
\gamma(k)=-H(0)+\sum_{n=1}^{\infty} c_n (k+1)^{2n},
\end{eqnarray}
at a finite $n$ (see the discussion of the truncation's consistency below). This truncation corresponds to the assumption that the cumulant series for $\gamma(k)$ given by Eq.~(\ref{gamma}) converges fast. This is the case if the correlation time of $\xi(t)$ is small. The simplest approximation is quadratic which holds rigorously in the limit of zero correlation time, the so-called Kraichnan model \cite{review}, cf. \cite{quadratic}. The form of the quadratic approximation is fixed uniquely by the demands that $\gamma(0)=\gamma(-2)=0$ and $\gamma'(0)=\lambda_1$. We find $\gamma(k)=\lambda_1 k(k+2)/2$ which implies $H(0)=\lambda_1/2$. It seems reasonable that for the typical case where the dimensionless correlation time is of the order of one, the quadratic approximation is too restrictive. However, a quartic approximation would already work well in typical cases for not too large values of $k$ given by cf. Eq.~(\ref{gamma}) and below.

In contrast, in the higher-dimensional case, the symmetry $\gamma(k)=\gamma(-d-k)$ and $H(\lambda)=H(-\lambda)-\lambda d$ holds only for the time-reversible statistics, see remark after Eq.~(\ref{rgtr}). For time reversible statistics, the quadratic approximation, holding in the Kraichnan model, is $\gamma(k)=\lambda_1 k(k+d)/d$ which implies $H(0)=\lambda_1 d/4$. Inspection of the data of \cite{mj} for the motion of tracers in the NS turbulence demonstrates $H(0)$ is appreciably larger than $3\lambda_1/4$. This is reasonable because the NS flow is neither time-reversible nor short correlated.

We consider developing a fitting function for the NS flow aimed at describing observations similar to \cite{mj,Bagheri,mj12,bc}. The observations demonstrate that $H(\lambda)$ is a smooth convex function which indicates that its Legendre $\gamma(k)$ can be approximated by a low order polynomial reasonably well. We propose a fitting procedure that uses properties whose measurement does not demand accumulation of large amounts of data. These are quantities derived from the most probable events: $\lambda_1$, $\lambda_d$ and $\Delta$. In contrast, the measurement of $\gamma(k)$ generally demands rare events, see Eq.~(\ref{trabs}).

Quadratic approximation of the Kraichnan model $\gamma(k)=\lambda_1 k(k+d)/d$ is too restrictive (the consideration is performed in $d$ dimensions and $d=3$ must be set for the case of interest). It gives a symmetric function with respect to $k=-d/2$ with $\lambda_d=-\lambda_1$ and $\Delta=2\lambda_1/d$, besides the already described $H(0)=\lambda_1 d/4$. These symmetries are appreciably violated by the NS flow. The difference of $\lambda_3$ and $-\lambda_1$ is not so large: we have $\lambda_3\approx -5\lambda_1/4$, see \cite{review}. Larger difference holds for the already considered $H(0)=3\lambda_1 /4$. Thus we resort to higher order, quartic polynomial approximation (cubic polynomial approximation can be degenerate because time-reversibility can be violated only weakly, e. g. the difference between $\lambda_3$ and $-\lambda_1$ is not so large for the homogeneous turbulence). The form of this approximation is fixed uniquely by the demands that $\gamma'(0)=\lambda_1$, $\gamma''(0)=\Delta$, $\gamma'(-d)=\lambda_3$ and the vanishing of $\gamma(k)$ at $k=0$ and $k=-d$,
\begin{eqnarray}\label{gammak}&&\!\!\!\!\!\!\!\!\!
\gamma(k)=\frac{\lambda_1 k(k+d)((k-k_0)^2+a^2)}{(k_0^2+a^2)d}, \label{quar}
\end{eqnarray}
where $k_0$ and $a^2$ are constants. These are fixed by the constraints $\gamma'(k=-d)=\lambda_d$ and $\gamma''(0)=\Delta$ giving,
\begin{eqnarray}&&\!\!\!\!\!\!\!\!\!
\frac{d^2+2dk_0}{k_0^2+a^2}=\frac{|\lambda_d|-\lambda_1}{\lambda_1},\ \
\frac{1}{d}-\frac{ 2k_0}{k_0^2+a^2}=\frac{\Delta}{2\lambda_1}.
\end{eqnarray}
The condition that $\gamma(k)$ produced by Eq.~(\ref{quar}) is convex is found by demanding that $\gamma''(k)>0$. This gives the demand
that the discriminant of the quadratic form in,
\begin{eqnarray}&&\!\!\!\!\!\!\!\!\!
\gamma''(k)
=\frac{2\lambda_1(6k^2\!+\!3(d-2k_0)k\!+\!k_0^2\!+\!a^2\!-\!2dk_0)}{(k_0^2+a^2)d}.
\end{eqnarray}
is negative,
\begin{eqnarray}&&\!\!\!\!\!\!\!\!\!
12k_0^2-24a^2+12dk_0+9d^2<0. \label{nd}
\end{eqnarray}
If the condition above does not hold, then the problem at hand is strongly non-quartic and Eq.~(\ref{quar}) is an invalid fit (globally). The approximation given by Eq.~(\ref{quar}) treats the points $k=0$ and $k=-d$ asymmetrically: we fix $\gamma''(0)=\Delta$ but we do not impose a similar demand for $\gamma''(-d)$, we leave that value as a free parameter in our approach. This can be remedied by considering the fit by a polynomial of fifth order. However it seems by qualitative comparison with $H(\lambda)$ in \cite{mj} and by quantitative comparison with \cite{Bagheri} below that the mistake introduced by the quartic approximation would not be large.

We remark that in the processing of experimental data, it could be useful to take $\eta(t)\equiv \rho_1(t)/t$ and $\theta(t)\equiv -\rho_3(t)/t$ as independent random variables for the parameterization of the PDF $P(\{\rho_i\}, t)$ where $\sum_{i=1}^3\rho_i(t)=0$ (we consider the NS case $d=3$). The domain of definition of these variables is $\eta\geq 0$ and $\theta\geq 0$ and we use,
\begin{eqnarray}&&\!\!\!\!\!\!\!\!\!
P(\eta, \theta, t)\sim \exp\left(-t{\tilde H}\left(\eta,\theta\right)\right).
\end{eqnarray}
Here we do not distinguish the notation for the large deviations function from Eq.~(\ref{ld}). The function ${\tilde H}\left(\eta,\theta\right)$ is symmetric for time-reversible statistics. We have for $\gamma(k)$,
\begin{eqnarray}&&\!\!\!\!\!\!\!
\gamma(k)\!\sim \!\frac{1}{t}\ln \left\langle \E{kt\eta}\right\rangle,\ \ k>-1;\nonumber \\&&\!\!\!\!\!\!\!
\gamma(k)\!\sim \!\frac{1}{t}\ln \left\langle  \E{t\left((k+1)\theta-(k+2)\eta\right)}\right\rangle,\ \ -2\!<\!k\!<\!-1; \nonumber\\&&\!\!\!\!\!\!\!
\gamma(k)\!\sim \!\frac{1}{t}\ln \left\langle \E{-(k+3)\theta}\right\rangle,\ \ k<-2.
\end{eqnarray}
These formulas allow the derivation of $\gamma(k)$ from the generalized Lyapunov exponents as defined in \cite{mj}.

Next, we introduce a possible measure of time irreversibility of the NS statistics. We observe that for time reversible statistics $H(\lambda)=H(-\lambda)-\lambda d$ implies $H'(0)=-d/2$. Thus deviations from the last identity can be used for quantifying irreversibility of the Lagrangian trajectories of the incompressible NS flow. This quantity can be obtained from the data of any of the works \cite{mj,Bagheri,mj12,bc} since all these works, despite the differences in the definitions, have this quantity. The data of \cite{mj,mj12,bc} provide $H(\lambda)$ for $\lambda\geq 0$ and \cite{Bagheri} give the full $H(x)$. Here we use the full data of \cite{Bagheri} for testing the quartic polynomial fitting.

\begin{figure}
  \centerline{\includegraphics[width=9.5cm]{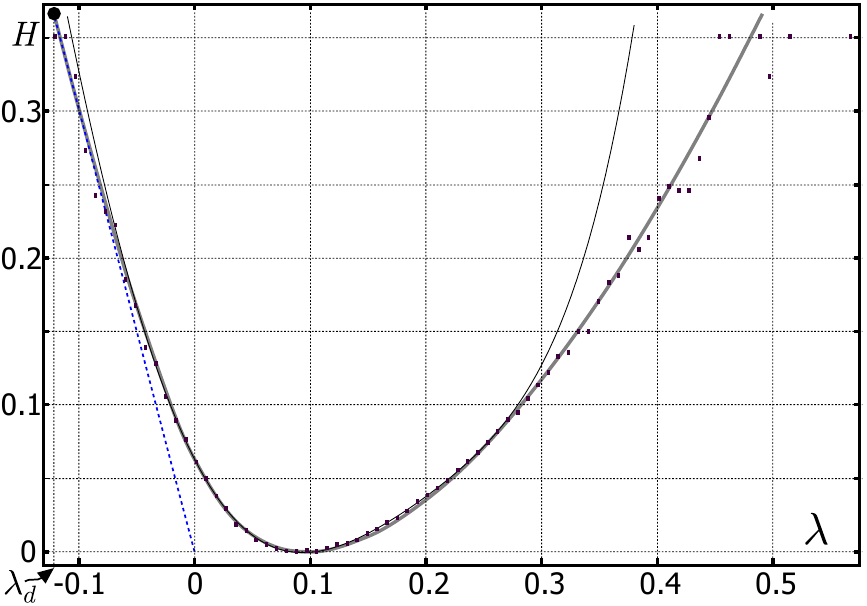}}
  \caption{Simulation results of \cite{Bagheri} for the large deviation function, courtesy of the authors, are shown as black dots. Direct fitting of $H(\lambda)$ with a quartic polynomial is shown as thin black curve. Much better results can be obtained if
the data are fitted to the Legendre transform of the quartic polynomial  of Eq. (\ref{gammak}), see the thick grey curve.
The intersection of the blue dotted line $H=-d\lambda$ with the $H(\lambda)$-curve at $\lambda=\lambda_d$ is shown as a black circle near the upper left corner of the graph.}
\label{Hla}
\end{figure}

\begin{figure}
  \centerline{\includegraphics[width=9.5cm]{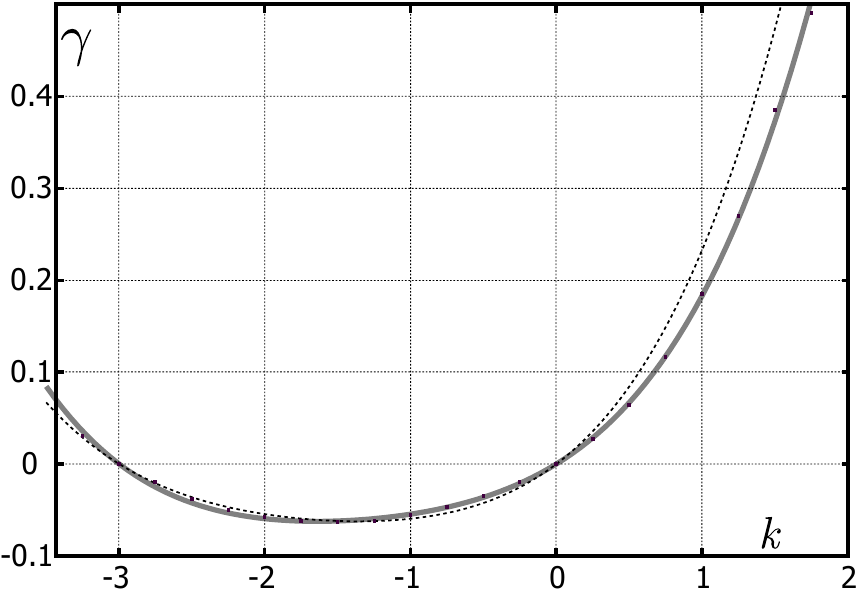}}
  \caption{Legendre transform of the data of Fig \ref{Hla} is shown as black dots; these data can be approximated with a quartic polynomial using Eq. (\ref{gammak}) with a very good accuracy, the corresponding least square fit is shown as thick grey curve. The dotted curve shows the mirror-symmetric polynomial $\tilde\gamma=\gamma(-3-k)$ and demonstrates the time-irreversibility of the flow: in the case of time-reversibility, $\gamma(-3-k)=\gamma(k)$.}
\label{Gk}
\end{figure}

We demonstrate how Eq. (\ref{gammak}) can be used based on the data of Fig. 4 of Ref. \cite{Bagheri}. By courtesy of the authors, we obtained tabulated value pairs of the large deviation function $H(\lambda)$ as a function of $\lambda$ (in  Ref. \cite{Bagheri} this function was denoted as $S(\mu)$). Our analysis here is based on the data points representing the results of the longest available simulation run (with the smallest statistical variance of the data) shown in Fig.~\ref{Hla} as black dots. As the first step, the value pairs  of $H(\lambda)$ were converted into tabulated value pairs of $\gamma(k)$ via Legendre transform. In order to minimize statistical fluctuations, the Legendre transform was applied to a cubic polymonial approximating
$k\lambda-H(\lambda)$ near its local maximum for each value of $k$ (the cubic polynomial was obtained via a least square fit using ten data points  in the neighbourhood of the local maximum).  The resulting data points were
approximated with a quartic  polynomial according to Eq. (\ref{gammak}) using least square fit. This produced the convex function $\gamma(k)=0.00390916k(k+3)(6.94324+(k+1.18176)^2)$ obeying Eq.~(\ref{nd}), see Fig.  \ref{Gk}. Based on this polynomial approximation, we can obtain the values of the Lyapunov exponents: $\lambda_1=0.0978$,  $\lambda_3=-0.1202$, and  $\lambda_2=0.0224$. This yields the exponent ratio $\lambda_3/\lambda_1=1.23$, which is essentially the same value as was
obtained in Ref. \cite{mj} suggesting that the dependnece of the Lyapunov exponent ratios is insensitive with respect to the
Reynolds number (while the data of Ref. \cite{mj} correspond to $\mathrm{Re}_\tau=1000$,
the data of \cite{Bagheri} used for the current calculations are based on $\mathrm{Re}_\tau=180$).

Finally, Legendre transform was applied to this quartic polynomial, to obtain a function approximating the large deviation function, shown as a grey thick line in Fig. \ref{Hla}. Note that the left tail of this curve is asymptotically linear while convergence to the asymptotic behaviour $H \propto  \lambda^{4/3}$ of the right tail is very slow and cannot be observed in this plot.
When comparing this Legendre-transform-aided quartic fit with the direct quartic fit (thin black line), it should be emphasized that while  the former has three fitting parameters (the values at $k=0$ and $k=3$ are fixed to $\gamma=0$), the latter has five fitting parameters. Our fit works well over the entire range, and hence, can be extrapolated towards the extreme deviations for which direct statistical data are usually inadequate.

We observe that our quartic polynomial fit, given by Eq.~(\ref{quar}) constrained by Eq.~(\ref{nd}), produces a valid probability density function described by the large deviations function $H(x)$ that obeys all the conditions necessary for the statistical realizability. However, this result might seem to contradict the Marcinkiewicz theorem that tells that there can be no statistics where all cumulants starting from some order larger than two vanish \cite{lu}. The resolution of this seeming contradiction is that we only describe the leading order term at large times. The cumulant generating function $\ln \left\langle \exp(k\rho(t))\right\rangle$ is not fit by a polynomial, rather it is given by $\gamma(k)t+o(t)$ where the correction terms will provide finite cumulants at any time $t$. The detailed study of how this situation would not produce a violation of the ridge inequality used in the theorem's proof \cite{lu} is beyond our scope here. It certainly provides an interesting question in the theory of characteristic functions. We confine ourselves with the demonstrated realizability of our fitting.

\section{Correlation dimension as zero of the generalized Lyapunov exponent} \label{cor}

In this Section we consider compressible flows. We study the correlation dimension $D$ of the multifractal support of the steady state density. This density is the random flow counterpart of the SRB measures \cite{ruelle}.
It was demonstrated in \cite{do,krzysztof} that $\gamma(-D)=0$ which we use here for finding approximations of $D$.

We observe that due to the convexity and positivity of $\gamma'(0)=\lambda_1$ (we consider a chaotic system with positive Lyapunov exponent), $\gamma(k)$ has only two zeros. Combining this with $\gamma(-d)>0$, see Sec.~\ref{ineq}, we find that the non-trivial zero of $\gamma(k)$ is located between $-d$ and $0$. This is necessary for consistency with $\gamma(-D)=0$ since a fractal dimension must be enclosed between zero and the dimension of space. We remark that in the case of $\lambda_1<0$ which could occur for some maps the attractor degenerates in points implying zero fractal dimension $D$ and coincidence of the two zeros of $\gamma(k)$.

\begin{figure}
  \centerline{\includegraphics[width=9.5cm]{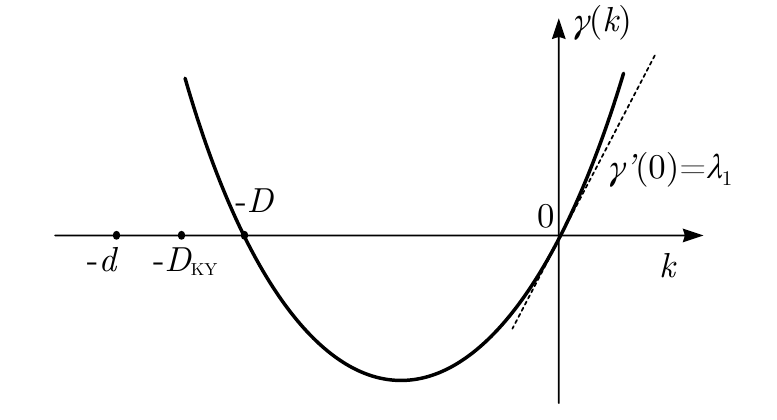}}
  \caption{Typical form of $\gamma(k)$ for compressible flows. The non-trivial zero is located at minus correlation dimension of the attractor and $\gamma$ at minus the Kaplan-Yorke dimension of the system is positive.}
\label{fig:k1}
\end{figure}

The understanding that the second zero of $\gamma(k)$ is located at minus the correlation dimension of the attractor brings a stronger result than $\gamma(-d)>0$. Correlation dimension is not larger than the information dimension \cite{hp} which is given by the famous Kaplan-Yorke (KY) \cite{ky} formula $D_{KY}=n+\delta$, where the integer $n$ and fractional dimension $0<\delta\leq 1$ are determined from the condition $\lambda_1+\ldots+\lambda_{n}+\delta \lambda_{n+1}=0$. Despite that counterexamples where KY formula does not hold can be constructed, typically the formula works and for random flows it can be proved \cite{ly}. We conclude from $D_{KY}>D$ and $\gamma(-D)=0$ that $\gamma(-D_{KY})>0$. Since $D_{KY}<d$, this result implies $\gamma(-d)>0$. The general form of $\gamma(k)$ is provided in Fig.~\ref{fig:k1}.

Development of polynomial approximations for $\gamma(k)$ is more difficult in the compressible case. This is because the incompressible flow conditions $\gamma(-d)=0$ and $\gamma'(-d)=\lambda_d$ are no longer true and have no simple counterparts in compressible flow, cf. Eqs.~(\ref{derivatives})-(\ref{identityp}). Thus, the position of the non-trivial zero of $\gamma(k)$ is no longer fixed; now it is positioned at $-D$, where the correlation dimension $D$ can take any value between zero and $d$.
 What we have instead are the two conditions $\gamma(-d, 1)=\gamma(-d, 2)=0$ described after Eq.~(\ref{identityp}). Similarly, instead of $\gamma'(-d)=\lambda_d$ we have the two conditions in Eq.~(\ref{derivatives}). We see that the polynomial approximation must be worked out for the full function $\gamma(k, p)$ and only then can $\gamma(k)$ be obtained as $\gamma(k, 0)$. The quadratic approximation, that holds in the limit of the small correlation time, reads \cite{thesis}
\begin{eqnarray}&&\!\!\!\!\!\!\!\!\!\!\!\!\!\!
\gamma(k, p)=\lambda_1 k+p\sum_{i=1}^d\lambda_i-p^2\sum_{i=1}^d\lambda_i
\nonumber\\&&\!\!\!\!\!\!\!\!\!\!\!\!\!\!
+\frac{k^2}{d}\left(\lambda_1-\frac{2\sum_{i=1}^d\lambda_i}{d}\right)-\frac{2\sum_{i=1}^d\lambda_i}{d}kp.
\end{eqnarray}
This form is uniquely fixed by the conditions $\gamma(0, 0)=\gamma(0, 1)=\gamma(-d, 1)=\gamma(-d, 2)=0$, $\partial_k \gamma(0, 0)=\lambda_1$, and $\partial_p \gamma(0, 0)=\sum_{i=1}^d\lambda_i$.
This formula can be proved in the Kraichnan model \cite{review}. We find from the equation above that $\gamma(k)=\gamma(k, 0)$ is given by
\begin{eqnarray}&&\!\!\!\!\!\!\!\!\!\!\!\!\!\!
\gamma(k)=\lambda_1 k+\frac{k^2}{d}\left(\lambda_1-\frac{2\sum_{i=1}^d\lambda_i}{d}\right),
\end{eqnarray}
which reduces to the previous formula for incompressible flow after setting $\sum_{i=1}^d\lambda_i=0$. Seemingly, this form could not be guessed without the preliminary description of $\gamma(k, p)$. We find that in this approximation, the correlation dimension, fixed from $\gamma(-D)=0$, is
\begin{eqnarray}&&\!\!\!\!\!\!\!\!\!\!\!\!\!\!
D=\frac{d}{1-2\sum_{i=1}^d\lambda_i/(\lambda_1 d)}.
\end{eqnarray}
Due to $\sum_{i=1}^d\lambda_i<0$, the equation provides $D$ which is smaller than $d$. This formula has seemingly not been proposed before. Despite its crudeness, it only gives twenty per cent mistake ($0.67$ versus the observed $0.86$) for $D$ in the strongly multifractal situation of tracers on a surface flow \cite{uw2} where the correlation time is not short \cite{bof}. In the small compressibility limit $|\sum_{i=1}^d\lambda_i/\lambda_1|\ll 1$ we find
\begin{eqnarray}&&\!\!\!\!\!\!\!\!\!\!\!\!\!\!
d-D\approx \frac{2|\sum_{i=1}^d\lambda_i|}{\lambda_1}. \label{cs}
\end{eqnarray}
In the case of short correlation time the statistics is time-reversible and $\lambda_1\approx-\lambda_d$ holds. Using the equality Eq.~(\ref{cs}) reduces to the known universal formula for the correlation dimension in the small compressibility limit that holds irrespective of the smallness of the correlation time, see \cite{FFS,fouxon1} and the next Section.

Our study of the incompressible case indicates that the quadratic approximation would usually be too restrictive however the quartic polynomial fit of $\gamma(k, p)$ would in many cases be very efficient. This fit can then be used for finding $D$ as the unique non-trivial solution of the quartic equation $\gamma(-D, 0)=0$. We consider the construction of the approximation. The quartic polynomial has fifteen unknown coefficients. We impose the conditions $\gamma(0, 0)=\gamma(0, 1)=0$ and $\gamma(0, p)=\gamma(-d, p+1)$. The last condition gives five constraints demanding equality of two polynomials of fourth order (it implies $\gamma(-d, 1)=\gamma(-d, 2)=0$).
Then we have the three conditions given by Eq.~(\ref{derivatives}). We have three more conditions which are besides $\partial_k \gamma(0, 0)=\lambda_1$ and $\partial_p \gamma(0, 0)=\sum_{i=1}^d\lambda_i$ also
$\partial_p \gamma(0, 1)=-\sum_{i=1}^d\lambda_i^-$. Finally the usage of dispersions $\partial_k^2 \gamma(0, 0)=\Delta$ and $\partial_p^2 \gamma(0, 0)=\Delta'$ allows to fix all the fifteen coefficients. The
coefficients must obey the realizability condition of positive Hessian of $\gamma(k, p)$. The resulting formulas are quite cumbersome and can be worked out separately in different practical cases of interest. Below we provide a different scheme of approximating the correlation dimension that might
provide a shortcut in some situations.

\section{Correlation dimension in the limit of small compressibility}

In this Section we consider the case where the correlation dimension is close to the space dimension. For incompressible flow $D=d$, and therefore, $D$ is close to $d$ when the compressibility of the particles' flow is small.
The location of the non-trivial zero of $\gamma(k)$ can be obtained with good accuracy by studying the Taylor expansion of $\gamma(k)$ near $k=-d$.

We derive the leading order approximation for $\gamma(-d)$ at small compressibility. We observe that at small compressibility, the flow divergence $w\equiv \nabla\cdot\bm v$ is small so the cumulant expansion of
\begin{eqnarray}&&\!\!\!\!\!\!\!\!\!\!\!\!\!\!
s(p)=\lim_{t\to\infty}\frac{1}{t}\ln\left\langle \exp\left(p\int_0^t  w(t', \bm q(t', \bm x)) \D t'\right)\right\rangle,
\end{eqnarray}
demonstrates that $s(p)$ is a parabola, cf. \cite{fouxon1}. Then the results of Sec. \ref{s} uniquely fix the form of $s(p)$ as $p(1-p)\sum_{i=1}^d \lambda_i$. We find using $\gamma(-d)=s(-1)$, see Sec.~\ref{ineq}, that
\begin{eqnarray}&&
\gamma(-d)=-2\sum_{i=1}^d \lambda_i>0.
\end{eqnarray}
This gives the leading order approximation for $\gamma(-d)$ at small compressibility (the zero order approximation here is zero). The leading order approximation for $\gamma'(-d)$ is its value for incompressible flow $\lambda_d$. We find that $\gamma(k)\approx -2\sum_{i=1}^d \lambda_i+\lambda_d(k+d)$ at $k+d\ll 1$. We conclude that the position of the non-trivial zero of $\gamma(k)$ in the leading order in small compressibility obeys
\begin{eqnarray}&&
d-D\approx \frac{2\sum_{i=1}^d \lambda_i}{\lambda_d}. \label{cord}
\end{eqnarray}
This coincides with Eq.~(\ref{cs}) on setting $\lambda_d=-\lambda_1$. It is readily seen from the definition \cite{ky} of $D_{KY}$ that at small compressibilty, $d-D_{KY}=\sum_{i=1}^d \lambda_i/\lambda_d$. We conclude that the correlation codimension $d-D$ is twice the Kaplan-Yorke codimension
$d-D_{KY}$. This result was found in \cite{FFS}, see also \cite{fouxon1}. Its self-consistency demands that $d-D\ll 1$. However the recent experimental confirmation of the
formula by \cite{fll} demonstrated that the Eq.~(\ref{cord}) can hold also when compressibility is already not very small and $d-D\approx 0.6$.

\section{Conclusions}

In this work, we derived properties of the generalized Lyapunov exponent $\gamma(k)$. We demonstrated that its study for compressible flows demands the introduction of a more general rate function $\gamma(k, p)$ that describes the joint growth rates of infinitesimal distances and volumes. The number of provided properties allows to fix the form of polynomial approximations for $\gamma(k, p)$ of up to fourth order. The approximation then gives $\gamma(k)$ as $\gamma(k, 0)$ and the correlation dimension as non-trivial solution of $\gamma(-D, 0)=0$. We derived the simplest quadratic approximation that holds if the correlation time is short. We demonstrated that its use beyond this range of validity still produces a good approximation for the correlation dimension in the case of the surface flow of tracers. However, generally this approximation is too restrictive. In contrast, the quartic polynomial approximation seems to be flexible for providing rather accurate approximations. Thus, we demonstrated that application of the proposed procedure to the incompressible turbulent channel flow provides nearly perfect fit for the numerical data of \cite{Bagheri}. It is plausible that the quartic polynomial approximation will work rather accurately since all observations known to us provide quite smooth $\gamma(k)$ that seem to be fittable
by the quartic polynomial. The proposed approximation can be particularly useful in complex situations when it is impossible to obtain large amounts of data:   $\gamma(k)$ and the correlation dimension can be found on the basis of a small number of measurements, it is sufficient to know the mean and the dispersion of the finite time Lyapunov exponents.
We also derived symmetry relations of the Gallavotti-Cohen type \cite{dor} which hold when there is statistical time-reversibility.

We have provided a different approximation scheme for the correlation dimension. This involves a formal expansion in the flow compressibility as a small parameter. This scheme is useful since weak compressibility occurs often in fluid mechanical applications. We demonstrated that the lowest order approximation reproduces other known results that are derived differently \cite{FFS}. This result was demonstrated experimentally to hold at not too small compressibility \cite{fll}. The approach proposed here, in contrast to the previous one, provides a route to the higher order approximations.

It must be stressed that despite that, we provided reasons why the quadratic approximation \cite{quadratic} for $\gamma(k)$ and $\gamma(k, p)$ is too restrictive, however sometimes it still applies. This is the case of water droplets
sedimenting in turbulence of cloud air, relevant for the rain formation problem \cite{FFS}. It was demonstrated in \cite{2015} that in the fast sedimentation limit, droplets can be described by a smooth spatial flow despite their strong inertia. This flow is short-correlated so that the quadratic approximation applies. This case is also characterized by small compressibility.

There is an intriguing question of the possible relation between the generalized dimensions and the fractal dimensions of the level sets of the first Lyapunov exponent. We consider the Lyapunov exponent's limit as a function of initial position of the pair,
\begin{eqnarray}&&
\lambda(\bm x)\equiv \lim_{t\to\infty} \frac{1}{t}\ln|W(t, \bm x){\hat r}|.
\end{eqnarray}
The function $\lambda(\bm x)$ is a constant, given by $\lambda_1$, for all $\bm x$ except for $\bm x$ whose total volume is zero (strictly speaking the Oseledec theorem asserts constancy on the set of full measure however generalization to the full volume can be done \cite{gawedzki}). The level sets $\lambda(\bm x)=\lambda\neq \lambda_1$ are fractals with certain Hausdorff dimension $d(\lambda)$, see \cite{yb}. We see that $H(\lambda)$, that describes the rate of disappearance of points with $t^{-1}\ln|W(t, \bm x){\hat r}|=\lambda\neq \lambda_1$, is quite similar to $d(\lambda)$. However these functions have different dimension and if a relation exists then it must involve a certain rate. The research of arising questions is left for future work.

We observe that our quartic polynomial fit for $\gamma(k)$ provides a way for addressing the dependence of the large deviations function of the NS turbulence on the Reynolds number. It is seen from the data of \cite{mj12} that this dependence is strong everywhere besides the left tail. This dependence can be studied by considering the Reynolds number dependence of the three parameters of our fit: $\lambda_1$, $\lambda_d$ and $\Delta$. This question is left for future work.

The approximation schemes developed here provide an efficient way for estimating correlation dimension of a chaotic attractor. Since this quantity has many applications, including collision kernel of particles, then we hope that the proposed scheme will find many uses in the future.

\section{Acknowledgements}

We are grateful to the authors of \cite{Bagheri} for providing the numerical data and specially to Dhrubaditya Mitra who helped in the data retrieval. We acknowledge the financial help of the Tallinn University of Technology.

{}

\end{document}